\documentclass[superscriptaddress,prd,aps,preprint,eqsecnum,floats]{revtex4}
\usepackage{graphicx}
\usepackage{epsfig}
\usepackage{amsfonts}
\usepackage{amssymb}
\usepackage{amsmath}
\usepackage{bm}
\usepackage{bbold}
  \newcommand\beq{\begin{equation}}
  \newcommand\eeq{\end{equation}}
  \newcommand\beqn{\begin{eqnarray}}
  \newcommand\eeqn{\end{eqnarray}}

   \newcommand {\nc} {\newcommand}

  \nc {\f} {\frac}                \nc {\s} {\sqrt}
   \nc {\gap} {\\[1ex]}
  \nc {\kap} {\kappa}		
  \nc {\amp}[1] {\phi_{#1}}
  \nc {\sgmtot} {\sigma_{\rm tot}}
  \nc {\aml}[2] {\phi_{#1}^{\mathrm{#2}}}
  \nc {\delsigL} {\Delta\sigma_{_{\mathrm L}}}  
  \nc {\delsigT} {\Delta\sigma_{_{\mathrm T}}}  

\def\pnul{\raise-.3ex\hbox{$\stackrel{\circ}{p}$}}\relax
\def\snul{\raise-.3ex\hbox{$\stackrel{\circ}{s}$}}\relax
\relax

\begin{document}
\title{\begin{flushright}{\small BNL-HET-07/14\\ Nov. 28, 2007 \\ \vspace{1 in} \endflushright}\begin{center} \bf{Spin asymmetries  for elastic proton scattering and \\ the spin dependent couplings of the Pomeron }
 \end{center}} 
\normalsize
\author{\normalsize T. L. Trueman} 
\affiliation{Physics Department, Brookhaven National Laboratory,
Upton, NY 11973}
\thanks{This manuscript has been authored under contract  with the U.S. Department of Energy. Accordingly, the U.S.
Government retains a non-exclusive, royalty-free license to publish or
reproduce the published form of this contributions, or allow others to do so,
for U.S. Government purposes.
\newpage
\thispagestyle{empty}
DISCLAIMER
This report was prepared as an account of work sponsored by an agency of the United States Government.  Neither the United States Government nor any agency thereof, nor any of their employees, nor any of their contractors, subcontractors, or their employees, makes any warranty, express or implied, or assumes any legal liability or responsibility for the accuracy, completeness, or any third partyÕs use or the results of such use of any information, apparatus, product, or process disclosed, or represents that its use would not infringe privately owned rights. Reference herein to any specific commercial product, process, or service by trade name, trademark, manufacturer, or otherwise, does not necessarily constitute or imply its endorsement, recommendation, or favoring by the United States Government or any agency thereof or its contractors or subcontractors.  The views and opinions of authors expressed herein do not necessarily state or reflect those of the United States Government or any agency thereof. 
}
\pagestyle{empty}
\begin{abstract}
This paper serves as a report on the large amount of analysis done in conjunction with the polarized proton program at RHIC. This comprises elastic scattering data of protons on protons in colliding beam or fixed target mode and proton beams on carbon targets. In addition to providing a model for the energy dependence of the  analyzing power of elastic scattering needed for proton polarimetry, it also provides some significant information about the spin dependence of dominant Regge poles. Most notably, the data indicates that the Pomeron has a significant spin-flip coupling. This allows the exploration of the double spin flip asymmetry $A_{NN}$ for which some data over a wide energy range is now available, along with a concrete realization of a proposed Odderon search. 
\end{abstract}
\maketitle
\section{Introduction}
\label{sec.1}
\setcounter{page}{1}
\pagestyle{plain}
Recent experiments carried out at RHIC using polarized proton beams give us the opportunity to address a question of long-standing interest; namely, what is the spin-dependence of the Regge couplings to the proton and, in particular, does the Pomeron coupling depend on the helicity of the proton? The work described in this paper results from a program of several years duration; background information and earlier results can be found in \cite{BKLST, RHICspin1,RHICspin2,RHICspin3, 2004}.
The program originated in the attempt to calculate the energy dependence of the analyzing power $A_N(t)$---the single transverse spin asymmetry---for the elastic scattering of polarized protons off protons or nuclei, especially carbon. This analyzing power can then be used for proton polarimetry in a high energy colliding beam facility for energies where no calibration standard exists or where rapid measurements are required. The goal of the present paper is to describe the physics that has been learned from this work. 

Starting from the well-justified assumption that high energy elastic proton-proton scattering can be described accurately with five Regge poles, the Pomeron and the dominant $C=\pm  1$ poles for $I=0,1$ \cite{Cudell}, we will use $A_N(t)$ measured at two energies in $p C$ scattering and at one energy in $pp$ scattering to determine the spin dependence of the five Regge residues in this description. As we will see, the analysis with this limited number of parameters requires a knowledge of the beam polarization $P$ at only one energy; at the other energy in $pC$ scattering, only the shape of the analyzing power in $t$ in the Coulomb Nuclear Interference region (CNI)  is required. 
The method makes use of the spin-dependent asymmetry $A_N(t)$ in the elastic scattering of two nucleons induced by the interference between the 1-photon exchange amplitude and the strong, hadronic scattering amplitude. The singularity of the  so-called Coulomb amplitude leads to a characteristic enhancement of $A_N(t)$  for very small $-t$ \cite{bigS, KL, BGL, BKLST}. This enhancement is important in yielding a practical signal. In order to separate the $I=0$ and $I=1$ exchanges for the Regge analysis it is necessary to have two targets; fortunately we have data for proton and carbon targets. Key to using $p p$ and $p C$ measurements together to extract physics (as opposed to simple polarimetry) is the relation, proved to be valid under a wide range of assumptions \cite{KT}, that the ratio of single-flip to nonflip amplitudes, commonly called $\tau$ (see Eq.~2.14 below), is the same for both processes; more precisely $\tau_{pC} =\tau_0\approx (\tau_{pp}  + \tau_{np})/2$.

Because these calculations were going on simultaneously with the experiments they are designed to describe, various ways of using the underlying Regge theory in order to extract the needed information from the limited data were tried \cite{RHICspin1, RHICspin2,RHICspin3,2004}.
Some of these will be described here because they shed some light on the validity of the as-yet-untested theory we are using; furthermore they may prove useful in other circumstances in the future.

This paper is organized in the following way. Section 2 will be a brief review of the general CNI theory used here, including quasi-elastic nuclear scattering. In Section 3.1 we will apply it to $pp$ data obtained at RHIC with proton beams at 24 {\rm GeV}/c and 100 {\rm GeV}/c on fixed proton targets \cite{pp100, pp24}. Errors will be estimated for the spin-flip factor $\tau(s)$. In Section 3.2 we will do the same thing for $pC$ RHIC  (and AGS) data, which  came chronologically mostly before the $pp$ data was obtained \cite{E950, osamu2004}. Because the target is a nucleus, the analysis is somewhat more complex than the $pp$ case and we will describe that here.  Errors  will be given here, too. In Section 4 we introduce the energy dependence using the Regge model of \cite{Cudell} and determine the energy dependence of the $pp$ flip factor $\tau(s)$ and the $pC$ flip factor $\tau_0(s)$ using several different methods. All of the Regge poles, in particular the Pomeron, are found to have significant spin-flip coupling. The $pC$ fit gives very good determination of the $I=0$ Regge couplings, including the Pomeron.  The $pp$ couplings are less certain. In Section 5 we will assume that the fit coefficients in \cite{Cudell} are given by factorizable Regge poles and determine the spin-flip couplings of the $\rho,\omega,f$, and $a_2$ . In Section 6 we will turn to the new and limited data on the double-spin asymmetry $A_{NN}$ \cite{okada2, wlodek2} and see what our model has to say about that. It will be necessary to take account of Regge cuts since factorization of pole couplings forces their contribution to $A_{NN}$ to vanish at $t=0$. This leads naturally to an exploration in Section 6.1  of the idea presented by Leader and the author \cite{LT} for searching for the Odderon. Finally Section 7 sums up.
\section{General CNI Discussion}
The method used here has a long history dating back at least to \cite{bigS}. In that paper, Schwinger proposed using the same effect to produce a beam of polarized neutrons. ({\it Almost } the same. The neutron singularity at $t=0$ is milder because it is neutral, but the magnetic moment provides the needed enhancement.) The details for $pp$ and $pC$ are slightly different and we summarize the needed formulas below. For an overview of polarization phenomena see \cite{bourrely}.
\subsection{Proton-proton CNI}
A detailed discussion can be found in \cite{BKLST}; a summary follows.
Five independent helicity amplitudes are required to describe
proton-proton elastic scattering \cite{BGL,GGMW} :
\begin{eqnarray}
\phi_1(s,t) & = & \langle ++|M|++ \rangle , \nonumber \\ \nonumber
\phi_2(s,t) & = & \langle ++ |M|-- \rangle , \\ \nonumber
\phi_3(s,t) & = & \langle +- |M|+- \rangle , \\  \nonumber
\phi_4(s,t) & = & \langle +- |M|-+ \rangle , \\ 
\phi_5(s,t) & = & \langle ++ |M|+- \rangle .
\end{eqnarray}
Here we use the normalization of \cite{BGL}. Since we are interested 
only in very high energy $\sqrt{s}$, such as will be
available at RHIC, and very small
momentum transfer
$|t| < 0.05 \mbox{ {\rm GeV}}^2$, we will generally neglect $m$ with respect to
$s$ and neglect $t$ with respect to $m$ to simplify the presentation of
the formulas which follow.  The total and differential cross sections are given by
\begin{equation} \label{eq:sigmatot}
\sgmtot = \frac{4\pi}{s}Im (\phi_1(s,t) + \phi_3(s,t))|_{t=0} 
\end{equation}
and
\begin{equation}
\frac{d\sigma}{dt} = \frac{2\pi}{s^2} \{|\phi_1|^2 + |\phi_2|^2 + |\phi_3|^2 +
|\phi_4|^2 + 4|\phi_5|^2\}.
\label{ppdiff}
\end{equation}
Using only initial state polarization, with one or 
both beams polarized,
one can measure seven spin dependent asymmetries. We follow the 
notation of \cite{BGL}. There are slight variations in the definitions used 
in the literature, having to do with the orientation of axes.
\begin{eqnarray} 
\label{eq:asymdef}
A_N \frac{d\sigma}{dt}& =& -\frac{4\pi}{s^2}  
Im \{\phi_5^*(\phi_1 +
\phi_2 +
\phi_3 -\phi_4)\}, \nonumber \\ A_{NN} \frac{d\sigma}{dt}& =& 
\frac{4\pi}{s^2}
\{2|\phi_5|^2 + Re (\phi_1^*
\phi_2 -
\phi_3^* \phi_4) \}, \nonumber \\ A_{SS}  \frac{d\sigma}{dt}& =& 
\frac{4 \pi}{s^2}
Re \{\phi_1 \phi_2^* + \phi_3
\phi_4^*\}, \nonumber \\ A_{SL} \frac{d\sigma}{dt}& =& \frac{4 \pi}{s^2}   
Re \{\phi_5^* (\phi_1 +
\phi_2  -
\phi_3+ \phi_4)\}, \nonumber \\ A_{LL}  \frac{d\sigma}{dt}& =& 
\frac{2 \pi}{s^2}
\{|\phi_1|^2 +|\phi_2|^2 -|\phi_3|^2 - |\phi_4|^2\}.  
\end{eqnarray}
It will be convenient to introduce some shorthand:
\beq		\label{eq.phi-plus}
 	 \phi_+  =  \case{1}{2}(\phi_1 + \phi_3) \, ,
\qquad
 	 \phi_-  =  \case{1}{2}(\phi_1 - \phi_3) \, ,
\eeq
which enter into
the two cross section differences corresponding to 
longitudinal and
transverse polarization:
\beq
	\f {\mathrm{Im}\,\phi_-(s,0)}{\mathrm{Im}\,\phi_+(s,0)}
	=					\phantom{-}
	\f{1}{2} \f{\delsigL(s)}{\sgmtot(s)} \, ,
\quad
	\delsigL = \sigma_{^\to_\gets} - \sigma_{^\to_\to},
 					\label{longitud}	
\eeq
\beq
	\f{\mathrm{Im}\, \phi_2(s,0)}{\mathrm{Im}\, \phi_+(s,0)}
	=
	- \phantom{\f{1}{2}} \f{\delsigT(s)}{\sgmtot(s)} \, ,
\quad
	\delsigT = \sigma_{_{\uparrow\downarrow}}
	-
	\sigma_{_{\uparrow \uparrow}}
 					\label{tranvers}.	
\eeq
At these small values of $t$, the interference of the strong amplitudes
with the single photon exchange amplitudes will be important; this interference 
is central
to this paper. To lowest order in $\alpha$, the fine structure constant,  one
replaces
\beq \amp{i} = 
        \amp{i}^{had} + \aml{i}{em}\exp(i\delta) G_E(q^2)^2
\eeq
	with hadronic and electromagnetic elements.
        The Coulomb or ``Bethe" phase $\delta $
is approximately independent of helicity \cite{BGL, cahn}
\beq
\delta = \alpha \ln\f{2}{q^2(B + 8/\Lambda^2)} - \alpha \gamma
\eeq
        where $B$ is the logarithmic 
derivative of the
differential cross section at $t=0$ and we use the fit
\beq
B(s)= 11 + 0.5 \ln{(s/10^2)}
\eeq
in ${\rm GeV}^{-2}$ through the RHIC region. Also $q^2=-t$, 
       Euler's constant $\gamma =
        0.5772\ldots$ and $\Lambda^2 = 0.71\mbox{ {\rm GeV}}^2$ reproduces
        the small momentum transfer dependence of the proton form
        factors assumed to satisfy
\beq
        G_E(q^2) = G_M(q^2)/\mu_p =(1 + q^2/\Lambda^2)^{-2}.
\eeq
For {\it pp} scattering at high $s$ and small $t$, the electromagnetic
	amplitudes are approximately
\beqn \label{eq:1 photon ex.}
	\aml{1}{em} &=& \aml{3}{em} = \f{\alpha s}{t} ,\nonumber
\gap
\aml{2}{em} &=& -\aml{4}{em} = \f{\alpha s \kappa^2}{4 m^2} , \nonumber
\gap
	\aml{5}{em} &=& -\f{\alpha s \kappa}{2 m\s{-t}},
\eeqn
	where $\mu_p = \kap + 1$ is the proton's magnetic moment,
	and $m$ its mass. For the full expressions see, e.g., \cite{BGL}. 
	For our purposes it will suffice to write
\beq
\phi_+(s,t)+\phi_-(s,t)= \frac{s}{4 \pi} \sgmtot(s) (i +\rho(s)) e^{B(s) t/2}
\eeq
where over the tiny CNI $t$-range we will neglect any variation of $\rho$. Likewise with
\beq
\phi_5(s,t)=\tau(s) \frac{\sqrt{-t}}{m} \phi_+(s,t)
\eeq
we will assume that $\tau$, which is complex, varies with $s$ but not with $t$ over the CNI range. (Sometimes in the literature one uses $r_5(s)= \phi_5(s,t) / (\sqrt{-t/m^2}\,{\rm Im} \phi_{+} )= \tau(s) (\rho(s) +i)$. We prefer to use $\tau$.) Then for simplicity for $pp$ scattering if we neglect $\delta$ and the amplitudes $\phi_2 \,, \phi_4$, which are not enhanced by the Coulomb singularity and are probably small (see Section 6 below),  the equation for $A_N$ in Eq.~(2.4) gives
\begin{widetext}
\beq
\frac{8 \pi m}{\sgmtot(s)^2 }
 \frac{d \sigma}{dt}e^{-Bt} \frac{A_N(s,t)}{\sqrt{-t}}
 = (\kappa/2 -{\rm Re}\,\tau(s)- \rho(s) {\rm Im\,}\tau(s)) \frac{t_c}{t} + {\rm Im\,}\tau(s) ( 1 + \rho(s)^2).
 \label{anpp}
\eeq
\end{widetext}
 $t_c=-8 \pi \alpha/\sigma_{tot}$.
This formula will be used with the recent data for $A_N(s,t)$ to determine the (complex) spin-flip factor for $pp$ scattering $\tau(s)$. 
\subsection{Proton-carbon CNI}
The basic physics here is the same as for $pp$, but it is simpler to describe because there are only two ampltudes   For $pC$ scattering the equation has the same form with all the quantities, including $t_c$ modified to refer to the carbon charge and wave function. Because the carbon is rather large, there will be significant variation with $t$ for  $\rho_{pC}$ and for the ratio of the electromagnetic to hadronic form factors. Also the Bethe phase is significant and needs to be kept. This was all worked out in \cite{KT} using the harmonic oscillator carbon wave function used by Glauber and Matthiae  \cite{GM, Glauber}. 
The needed functions are: \newline
single particle densities
\beqn
\rho_s(r)=2\,(\f{a_C}{\pi})^{\f{3}{2}} e^{-a_C\, r^2}, \rho_p(r)=\f{8}{3}\, a_C\, r^2 \, ( \f{a_C}{\pi})^{\f{3}{2}}   e^{-a_C\, r^2} \nonumber \\
\eeqn
the corresponding electromagnetic form factor
  \beqn
 F^{em}(t) = 4 \,a_C \int db J_0(\sqrt{-t} b) b\, e^{-a_C b^2} (1 + \f{4}{3}\,a_C(b^2 + \f{2}{ a_C})) \nonumber  \\
\label{em form}
\eeqn
the hadronic amplitude
\beqn
F_A^h(t)= {\rm Im} F^A_0(q) /{\rm Im} F^A_0(0) 
\eeqn
where for carbon  \cite{bassel}
\begin{widetext}
\beqn
F^C_0(q) &=&
i\,\int d^2b\, e^{i\vec q\cdot\vec b}\,
\left\{1 -\left[1-\frac{a_C \,\sigma_{tot} \,(1 -
i\,\rho)}{2\pi (1 + 2 B \, a_C)} \exp \left(-\frac{a_C \,b^2}{1 + 2
B \,a_C}  \right)\right]^4\right.
\nonumber\\ &\times&  \left.
\left[1 - \frac{a_C \,\sigma_{tot}\,
(1 - i \rho)}{2\pi (1 +2 B \,a_C)}\,
\left(1 -
\frac{2}{3(1 + 2 B \,a_C)} + \frac{2 a_C\,b^2}{3 (1 +2 B \,
a_C)^2}\right)\right.\right.
\nonumber\\ &\times& \left.\left.
\exp\left(-\frac{a_C\, b^2}
{1 + 2 B \,a_C}\right)\right]^8\right\}.
\label{hadronic}
 \eeqn
 \end{widetext}
 Note that $\rho$ in this expression denotes the $I=0$ nucleon-nucleon real-to-imaginary ratio, very close to the $pp$ value. $a_C=0.0152\, {\rm GeV}^2$.
The analog of Eq.~\ref{anpp} is
\begin{widetext}
\beqn
&& \frac{16\,\pi}{(\sigma^{pA}_{tot})^2}\,
\frac{d\,\sigma_{pA}}{d\,t}\,A^{pA}_N(t) =
\frac{\sqrt{-t}}{m_N}\,
F_A^h(t)\,\biggl\{F_A^{em}(t)\,\frac{t_c^A}{t}\,
\Bigl[\kappa \,(1-\delta_{pA}\,\rho_{pA})
\nonumber\\ \label{anpc}
&-&  2\,{\rm Re}\, \tau_0-(\delta_{pA}+ \rho_{pA})\,)\Bigr]
- 2\,F_A^h(t)\Bigl({\rm Im}\,\tau_0 (1 +
\rho_{pA}^2)\,\Bigr)\biggr\}.
\label{ancp}
 \eeqn
with 
\beqn
\frac{16\,\pi}{(\sigma^{pA}_{tot})^2}\,
\frac{d\,\sigma_{pA}}{d\,t} &=&
\left(\frac{t_c^A}{t}\right)^2\,\Bigl[F_A^{em}(t)\Bigr]^2 -
2\,(\rho_{pA}+\delta_{pA})\,\frac{t_c^A}{t}\,
F_A^h(t)\,F_A^{em}(t)\,\nonumber\\
&+& \left(1+\rho_{pA}^2-\frac{t}{m_p^2}\,|\tau_0|^2(1 + \rho_{pA}^2)\right)\,
\Bigl[F_A^h(t)\Bigr]^2
\label{padif}
 \eeqn
 \end{widetext}
 where
 \beq
 t_c^A=- 8 \pi Z \alpha / \sgmtot^{pA} .
 \eeq
 with
 \beq
\rho_{pA}(s,t)=Re{F^A_0(q)}/Im{F^A_0(q)}.
\eeq
\subsubsection{Inelastic corrections}
The above calculation was used in the paper with Kopeliovich \cite{KT} and all of my reports prior to 2004 \cite{RHICspin1, RHICspin2, RHICspin3}. Just after the last of these a preliminary measurement of the $pC$ differential cross section was reported. Although it is still preliminary, it is clear from the data that the differential cross section calculated in \cite{KT} does not agree with it. This becomes very clear above the CNI region where at diffraction dip predicted at $t=-0.09$ becomes in the data a break in the exponential fall which has a slope of about $60 \,{\rm GeV}^{-2}$ between -.02 and -.06 \cite{Bravar2} . Exactly this effect was reported long ago in the CERN proton-nucleus measurements \cite{Bellettini} where it was explained by additional incoherent proton-nucleon scattering which comes in because of the resolution. Subsequently, Glauber and Matthiae \cite{GM} extended the Glauber method used by \cite{KT} to take into account quasi-elastic scatterings.
 It is based on completeness and only requires that the experimental resolution be wide enough to include nearly all excited nuclear states. The mass resolution presented at Blois 2005 by Bravar is about 1 {\rm GeV} \cite{Bravar}. Since the binding energy of carbon is about 92 MeV all of the excited states that do not lead to production or breakup must be included and the completeness assumption is good. We will expand in powers of the scattering and take just the first term to estimate this correction. The key parameter is $N_1(s)$ which determines the number of quasi-elastic scattering. According to  \cite{GM} it is given by
\beq
N_1(s)= 2 \pi  \int  b\, db e^{-\sgmtot(s) T(b)}T(b)
\eeq
and the nuclear thickness function $T(b)$ is given by the integral over the nuclear density
\beq
T(b)=4 ( \f{ a_C}{\pi})^\f{3}{2}  \int^{\infty} _{-\infty }dx (1 + \f{4}{3}a_C (x^2 + b^2))e^{- (x^2 + b^2)a_C}.
\eeq
$N_1(s)$ varies slowly with $s$ and is approximately three quasi-elastic scatterings throughout the energy range.
The measured $pC$ cross section is then the sum of elastic cross section Eq.~\ref{padif} plus $N_1(s)$ times~Eq.\,\ref{ppdiff}.  The comparison between the elastic cross section and this sum is shown in Fig.~1.
\begin{figure}[htbp]
{\includegraphics{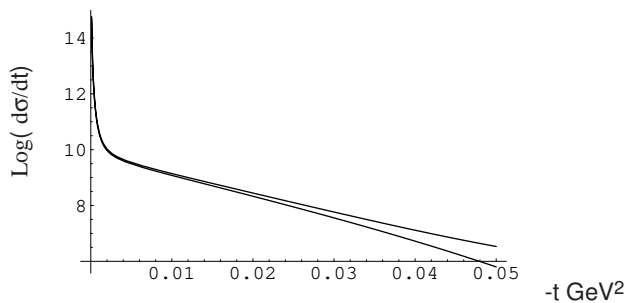}}	
\caption{$ \ln{\frac{d \sigma}{dt}}$ at $p_L=24 \, {\rm GeV}/c $. 
The lower curve is from \cite{KT}, the upper curve is based on \cite{GM} and includes quasi-elastic corrections}
\label{1}
\end{figure}
The difference is not large, but it is significant at the largest $-t$. For completeness Fig.~2 shows a larger range including the diffraction dip. An exponential fit to this calculation between -.02 and -.06 gives a slope of 58.2 ${\rm GeV}^{-2}$, close to the CERN slope \cite{Bellettini}.
\begin{figure}[htbp]
\includegraphics{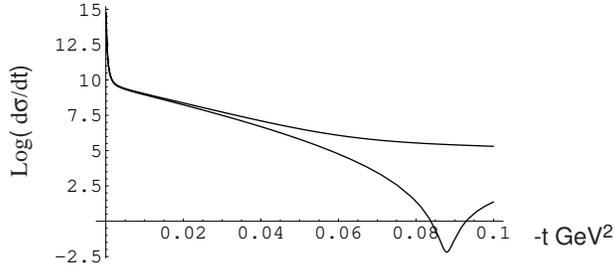}																\caption{$ \ln{\frac{d \sigma}{dt}}$ at $p_L=100 \, {\rm GeV}/c $ over a larger range of $t$.
The lower curve is from \cite{KT}, the upper curve is based on \cite{GM} and includes quasi-elastic corrections}
\label{2}
\end{figure}
\section{spin flip factors}
\subsection{pp}
$A_N(s,t)$ has been measured at RHIC for proton beam energies  of $24\,{\rm GeV}/c$ and $100 \,{\rm GeV}/c$ on a polarized gas jet target \cite{gasjet}. The data has been fit to the formula given in Section 2 and the results for $r_5$ reported \cite{pp100}, \cite{pp24}. We have converted these to our preferred parameter $\tau(s)=r_5(s)/(i + \rho(s)) $:
\beqn
\tau(s[100])&=&-0.0148 +0.002 i, \\
\tau(s[24])&=&-0.100 +0.0306 i
\eeqn
$s[p]=  2 m^2 + 2 m \sqrt{p^2 + m^2}$. (Because it occurs so often, we will often write $\tau(p)$ instead of $\tau(s[p])$ when there is no chance of confusion.) The quoted chi-squares for these determinations are: at 100 GeV =11.1 for 14 points and at 24 GeV = 2.87 for 9 points.
The error matrices were also given and used in our calculation of the $1\sigma$ error ellipses, shown in~Fig.3 and~Fig.4.
\begin{figure}[htbp]
\includegraphics{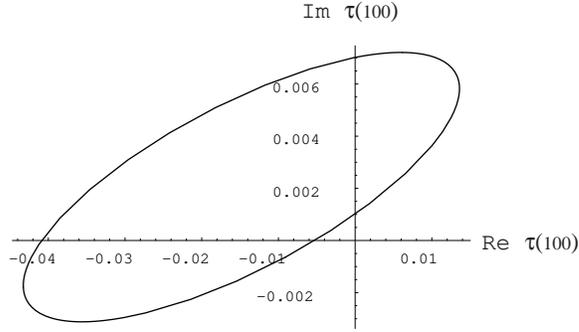}																\caption{$1\sigma$ error ellipse for the proton $\tau$ determined with $p_L=100\,{\rm GeV}/c$ beam on a polarized gas-jet target.}
\label{3}
\end{figure}
\begin{figure}[htbp]
\includegraphics{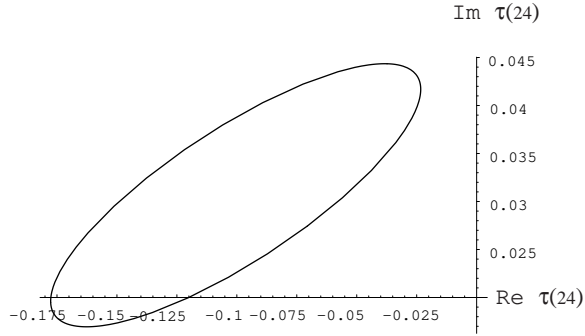}																\caption{$1\sigma$ error ellipse for the proton $\tau$ determined with $p_L=24\, {\rm GeV}/c$ beam on a polarized gas-jet target.}
\label{4}
\end{figure}
The best fit at 100\,{\rm GeV}/c has $\tau$ consistent with 0, as reported by the experimentalists; the error ellipse for the much shorter run at 24\,{\rm GeV}/c is about 3 times larger than the  100\,{\rm GeV}/c case but the value of $\tau$ is significantly non-zero. This was the last data to become available and had a decisive influence  on our results. The errors will surely be improved in the future.
\subsection{pC}
This process has a longer and more complex history than that of $pp$. The first results come from an AGS experiment E950 \cite{E950} at 21.7\,{\rm GeV}/c. There were also measurements at RHIC at 24\,{\rm GeV}/c and 100 \,{\rm GeV}/c.  The beam in the E950 experiment had its polarization determined by a completely independent experiment E925 based on a known analyzing power for $pp$ scattering \cite{E925}. The value they used was $0.407 \pm 0.036 (stat) \pm  0.049 (syst)$. For a few years the RHIC measurements  above injection energy used only estimates until 2004 when the polarized jet target became available at 100 {\rm GeV}/c. At that point a really extensive set of data with very small errors was obtained. In general, the experimentalists analyzed their data using~Eq.\ref{anpc} and reported $\tau$ values. (In fact, for historical reasons, they usually report values for $ r_{5\,pC}= \tau_0 (\rho_{pC} + i)$. Here we will always convert those numbers to $\tau$.) 
Because we now know that there are corrections due to quasi-elastic scattering, we have reanalyzed their data using the corrected form for the asymmetry: i.e. for each polarization state of the proton beam we calculate the sum of the corresponding elastic $pC$ differential cross section plus the $N_1$ times the $pp$ differential cross section. The difference of these two divided by the sum is the corrected analyzing power as a function of $\tau_0$. The value of $\tau_0$ determined in this way is a little different from the uncorrected fit from E950 \cite{E950}, but the chi-square is very much better, 0.55/d.o.f. compared to 1.16/d.o.f . The same comparison for the 100 GeV/c data gives 1.32/d.o.f. for the corrected fit compared to 1.51/d.o.f for the uncorrected fit.  We will use the corrected fits for each of the data sets we use. The results for E950 and for RHIC at 100 GeV/c are
\beqn
\tau_0(21.7)= -0.222 -0.0584i \\
\tau_0(100)=-.011 -.0498 i
\eeqn
The subscript 0 indicates that this is the $\tau$ value for $I=0$ exchanges. (At 100 {\rm GeV}/c we have used only the lowest 10 values of $-t$, omitting the last 4 points they report. The chi-square is considerably better that way and the corrections to the low $t$ approximations used in deriving Eq.~\ref{anpc} are smaller.)
The error ellipses are shown in In~Fig.5 and~Fig.6.
\begin{figure}[htbp]
\includegraphics{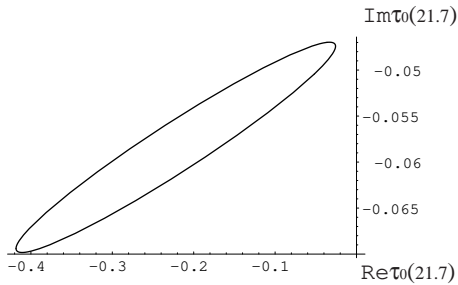}																\caption{$pC$ error ellipse for E950}
\label{5}
\end{figure}
\begin{figure}[htbp]
\includegraphics{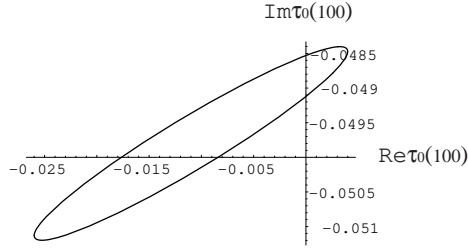}																\caption{$pC$ error ellipse for 100 {\rm GeV/c}}
\label{6}
\end{figure}
From these figures we easily see that the spin-flip factor is non-zero at both energies. Indeed, at 21.7 {\rm GeV}/c the flip factor is bigger than 20\%, and there is significant energy dependence.
The fits to the two data sets are shown in~Fig.7 and~Fig.8. (The data for E950 are from \cite{E950}. The  data at 100 {\rm GeV} are from \cite{osamu2004}). Notice how precise the latter set is. As a result it will play a central role in our parameter determination. The chi-square for 100 {\rm GeV} fit is 10.6 for 10 data points, significantly better than the best fit without quasi-elastic corrections which is  12.1 for 10 data points.
\begin{figure}[htbp]
\includegraphics{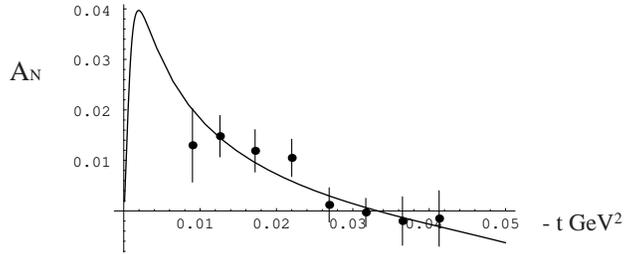}																\caption{$pC$ analyzing power for E950 \cite{E950}}
\label{7}
\end{figure}
\begin{figure}[htbp]
\includegraphics{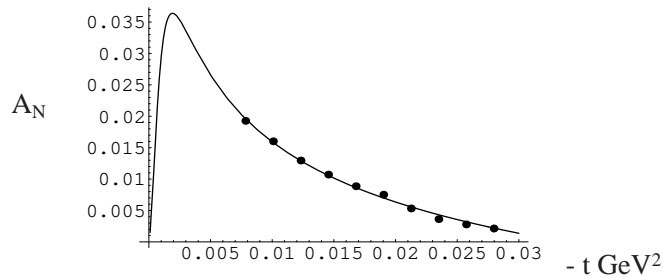}	
 \caption{$pC$ analyzing power for $p_L=100 \,{\rm GeV}/c$ with data from \cite{osamu2004}}
\label{8}
\end{figure}
\section{Energy dependence of the spin-flip factors $\tau(s)$}
It would be very useful to know the energy dependence of $\tau(s)$ so that the analyzing power could be used for polarimetry at energies where it has not been measured. At a deeper theoretical level,  it should give us some insight into the old  problem of the spin-dependence of Regge couplings \cite{MT, berger,irving}. We propose to take here a very simple approach which is appropriate for the energy and $t$ range of the physics under consideration. For an overview of Regge theory see \cite{pdb}.
From among all the various Regge fits, we have chosen to use the one  given by the Particle Data Group due to Cudell et al. \cite{Cudell}. This is based on a Regge picture with two simple Regge poles with $C=\pm1$ with $0 \le J \le 1$ at $t=0$ plus a more complex structure at $J=1$ representing the Pomeron. (In the earlier reports \cite{RHICspin1, RHICspin2, RHICspin3} we used an alternative fit from \cite{Cudell} in which the Pomeron was parametrized as a simple pole. This contributes to the difference between the spin-flip factors given here and those in the earlier reports.) The formula for the forward $pp$ amplitude 
\beqn
 \frac{8\pi}{s} \phi_+(s,0) &=&g_P(s) + g_{+}(s) + g_{-}(s) \\ \nonumber
 {}&=&g_0(s)
 \eeqn
 with 
 \beqn
g_P(s)&=&0.789 \,\pi  \ln{(s/29.1)} \\ \nonumber
& &+ i(91.26+ 0.789 \ln^2{(s/29.1)}), \\ \nonumber
g_{+}(s)&= &-Y_1\, s^{-\eta}( \cot\pi/2(1-\eta) -i), \\ \nonumber
g_{-}(s)&=&-Y_2 \,s^{-\eta '} (\tan\pi/2(1-\eta ') + i),
\label{ppel}
\eeqn
with $Y_1=109.51\, {\rm {\rm GeV}}^{-2}, Y_2=85.86 \,{\rm {\rm GeV}}^{-2}$.
The parameters are $\eta =0.458, \eta '=0.545$ and the coefficients are all in units of $\rm{{\rm GeV}}^{-2}$. $g_{+}$ and $g_{-}$ arise as sums of contributions of $C=+1$  ($f $ and $a2$) and $C=-1$ ($\omega$ and $\rho$) of $I=0$ and $I=1$, respectively, assumed in each case to be degenerate in $J$.

To introduce spin dependence we write
\beq
\ \frac{8\pi}{s} \phi_5(s,t) =\f{ \sqrt{-t}}{m} \{\tau_P \, g_P(s) + \tau_{+}g_{+}(s) + \tau_{-}g_{-}(s)\}
\label{phi5}
\eeq
For simple Regge poles $\tau_R$ is real and we will assume the same is true for the Pomeron.
Then 
\beq
\tau(s)=\f{\tau_P \, g_P(s) + \tau_{+}g_{+}(s) + \tau_{-}g_{-}(s)}{g_P(s) + g_{+}(s) + g_{-}(s)}
\label{tau (s)}
\eeq
and the energy dependence of the real and imaginary parts of $\tau$ are given by three real parameters.

The $pC$ scattering is pure $I=0$ and so requires a different set of functions: presumably, the Pomeron amplitudes are the same but the $C=+1$ amplitude is pure $f$ and the $C=-1$ amplitude is pure 
$\omega$ so $g_{+}$ and $g_{-}$ are replaced in the $pp$ amplitude to get the $pC$ amplitude by $g_f$ and $g_{\omega}$, respectively. From \cite{Cudell}
\beqn
g_{f}(s)&=& -Y_f\, s^{-\eta}( \cot\pi/2(1-\eta) -i) \\ \nonumber
g_{\omega}(s)&=&-Y_{\omega} \,s^{-\eta '} (\tan\pi/2(1-\eta ') + i).
\eeqn
with $Y_f=106.36\, {\rm GeV}^{-2}$ and $Y_{\omega}=81.49 \, {\rm GeV}^{-2}$,
not so different from Eq.~\ref{ppel}, because the  spin-independent parts of $pp$ and $pn$ scattering are nearly the same.

In obvious notation
\beq
\tau_0(s)=\f{\tau_P g_P(s) + \tau_{f}g_{f}(s) + \tau_{\omega}g_{\omega}(s)}{g_P(s) + g_{f}(s) + g_{\omega}(s)}.
\label{tau0(s)}
\eeq

\subsection{Determination of the $I=0$ spin-flip factors}
The measurement of the analyzing power at any energy $s$ determines two parameters, $\rm {Re}\,\tau(s)$ and $\rm {Im}\,\tau (s)$ and so knowledge of $A_N$ at two energies is more than enough to fix the individual spin-flip factors. Early in this work we encountered a situation where the {\em asymmetry} was measured but the polarization was not known. By separating the singular and non-singular terms in Eq.~\ref{anpc} one can determine the {\em shape} of the asymmetry  $S_0(s) $, in $t$. This is independent of the polarization $P$ and provides another relation between $\rm {Re}\,\tau$ and $\rm {Im}\,\tau$ at that energy:
\beq
S_0(s) \approx \f{P \,\rm{Im}\, \tau_0(s)}{P(\kappa/2 -\rm{Re}\,  \tau_0(s))}.
\eeq
Therefore knowledge of the analyzing power at one energy and the shape at another is sufficient to determine the flip factors and, thereby, the polarization/analyzing power at the second energy.

\subsubsection{E950 $\tau$ and the shape of the first 100 {\rm GeV} RHIC data:a test}
Historically, the first application of this method was to the first reported data (2002) for polarized protons on carbon at RHIC  \cite{dima}.  In addition to data with a proton beam whose polarization was measured using the E950 calibration, the asymmetry was also measured at 100 {\rm GeV} where the polarization was unknown. We fit data in Table 1 taken from \cite{dima},  
\begin{table}[h]
\centering
$\begin{array}{|c|c|c|}
\hline
\mbox{-$t$} & \mbox{$\epsilon(t)$}& \mbox{$e(t$)} \\ 
\hline
0.0117 & 0.0036 & 0.00055 \\
0.0138 & 0.0029 & 0.00047 \\
0.015 & 0.0034 & 0.00060 \\ 
0.0184 & 0.0018& 0.00069 \\ 
0.0194 & 0.0014 & 0.00058 \\
0.0217 & 0.0025  & 0.00087 \\
0.0249 & -0.0004 & 0.00072 \\ 
0.0251 & 0.0009 & 0.00085 \\
0.0306 & -0.0010 & 0.00072 \\
0.0360 & 0.0010 & 0.00091 \\
0.0416 & 0.0013 & 0.00116 \\
0.0473 & -0.003 & 0.00146
\\ \hline 
\label{dadada}
\end{array}$
\caption{\sl -t, raw asymmetry $\epsilon$$(t)$, and errors $e(t)$ for RHIC 100
{\rm GeV}/c}
\end{table}
to the formula in Eq.(\ref{anpc}) with the
right hand side multiplied by the unknown $P(100)$.  The result of the regression is 
 \begin{eqnarray}P(100)
(1- \frac{2}{\kappa} \rm{Re}\,\tau(100))&= &0.263  \\ \nonumber 
 P(100) \frac{2}{\kappa }\rm{Im}\,\tau(100)&=& -0.0137.
 \end{eqnarray}
 Combining these together we get for the shape of
the distribution
\begin{equation} \label{shape100}
S_0(100)= -\frac{0.0137}{0.263}=-0.052
\end{equation}
By expesssing the value of $\tau(21.7)$ as determined in E950 in terms of
the Regge spin-flip couplings, and in the same way express
$S_0(100)$ in terms of the Regge spin-flip couplings via $\tau_0(100)$, we have three equations to solve with the result
$\tau_{P{\rm test} }= 0.028 \pm 0.14 ,\tau_{f {\rm test} }= -0.967 \pm 0.35 ,\tau_{\omega{\rm test} } =0.509 \pm 0.23.$
There are significant errors in these determinations and we will move on to a more accurate determination in the next section. 

We could now calculate $\tau_{0{\rm test} }(s)$ at any higher energy, with corresponding accuracy.(The model as it stands, is not really suitable for going to lower energy because lower lying Regge poles will rapidly beome important. Thanks to Boris Kopeliovich for emphasizing this limitation \cite{KNP, Kramer, berger}.)  

To test and illustrate the model, we will use these results first for  $\tau_{0{\rm test}}(100)$ and thereby determine
the polarization at 100 {\rm GeV}/c, $P(100)$. 
Using these values of the Regge coupling we get
$\tau_0 (100) = -0.097 \pm 0.15  -( 0.054 \pm 0.02) i$.
 With errors this large this method will give only a crude indication of $P$. We find
$
P(100) = 0.23 \pm 0.08.
$
This is a little smaller, about 15\%, than the value 0.27 used by \cite{dima} based on assuming $A_N$ to be energy independent with the E950 value. In Fig.~9 we show the raw asymmetry measured at 100 {\rm GeV}/c plotted with the prediction using  Eq.~2.19  and  $P=0.23$.
The agreement is reasonable,  with $\chi^2$/dof about 1.5. 
\begin{figure}[thb]
\centerline{\epsfbox{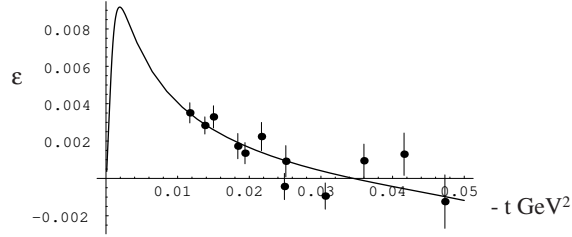}}
\medskip
{\caption[Delta]{\it Raw asymmetry $\epsilon$ at 100 {\rm GeV}/c \cite{dima} with curve predicted by the model $\tau_0(100)$
and polarization $P$, predicted to be 0.23}}
\label{asym100}
\end{figure}

\subsubsection{Shape of E950 and precision measurement at 100 {\rm GeV}}

The errors on $P$ and $\tau$ determined in the previous section are necessarily large because the errors on both data sets are large. We should be able to do much better by using 100 {\rm GeV} $\tau$ values determined in Section 3.2 which have very small errors and use just the shape from E950, which has a much smaller error than $\tau0(21.7)$ itself. The measured value of the shape is $S(21.7)=-0.0523$, using the quasielastic corrected $A_N$, and the error on it can be determined from the propagation of errors from the regression of  the 21.7 {\rm GeV} data. It is very small: $\delta \rm{Shape(21.7)}=0.003$. 

We can  use that along with the error matrix for $\tau(100)$ to determine the error matrices and ellipses  for the three Regge spin-flip couplings, Fig.~\ref{regge ellipses}:
\begin{figure*}[thb]
\centerline{\epsfbox{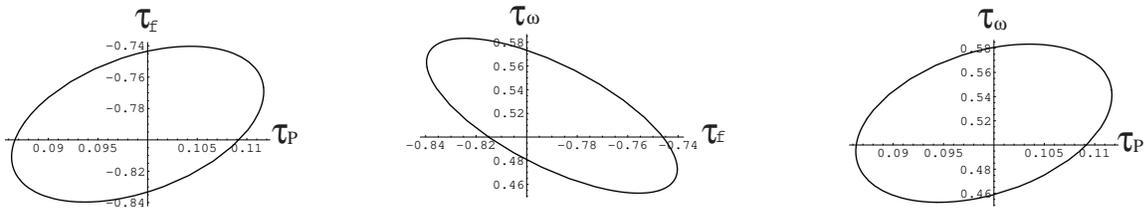}}
\medskip
{\caption[Delta]{\it Regge coupling error ellipses}
\label{regge ellipses}}
\end{figure*}
The errors are seen to range from about 6\% to 13\%.

 We have tried alternative determinations of $\tau_P$ using the 100 GeV/c data and the 21.7 GeV/c data in various ways with this model, but this one has the smallest errors, and it is certainly non-zero.
We will use the determination of residues here to calculate $\tau_0(s)$ in the following. The values are
\beqn
\tau_P=0.10 \pm 0.01, \,
\tau_f=-0.79 \pm  0.05, \,
\tau_{\omega}=0.52 \pm 0.06. \nonumber \\
\eeqn
Note that here $\tau_P$ is clearly non-zero, while in the ``test" case it is consistent with zero (as well as with this non-zero value.)

These give the energy dependence of $\tau_0$ as shown in Fig.~11.
\begin{figure}[thb]
\centerline{\epsfbox{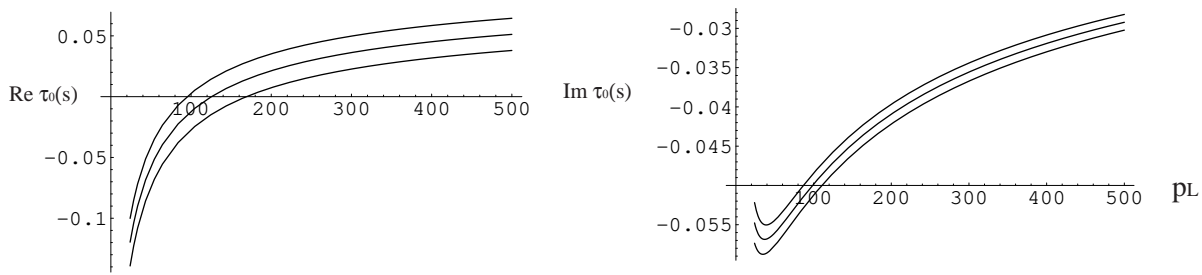}}   
\medskip
{\caption{\it The energy dependence of ${\mathrm Re}\,\tau_0$ and ${\mathrm Im}\,\tau_0$ through RHIC fixed target range with $1\sigma$ errors.}}
\label{tau0(s)}
\end{figure}

We can use the Regge coupling error matrices to determine the errors on the prediction for $\tau0(21.7)$, Fig.~12. The central value is consistent with the experimental measurement which has, for the real part, a much larger error than that of the prediction. 
\begin{figure}[h!]
\centerline{\epsfbox{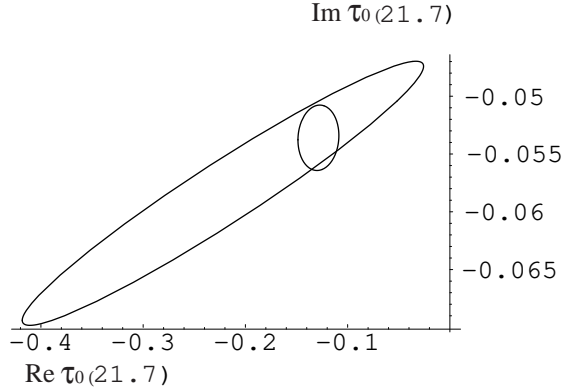}}
\medskip
{\caption[Delta]{Error ellipse for predicted $\tau(21.7)$ shown within the measured ellipse as determined by E950. See Fig.5 }
\label{ellipse21.7}}
\end{figure}

There also exists a set of $pC$ asymmetry $\epsilon$ data taken at 24 GeV/c at RHIC, but the final results (with errors) have not been released and so have not been used in the present analysis.
\subsection{Determination of the proton spin-flip factors}
We will describe here two distinct ways of determining the proton spin-flip factors. If the model and the data were perfect, they should give the same results.
\subsubsection{100 {\rm GeV}/c $pp$ measurement in conjunction with $\tau_P$ from $pC$}
The first $pp$ data that was available to us was the 100 {\rm GeV}/c data reported in Section~2. That provides us with two new parameters to determine the two $C=\pm 1$  spin-flip residues which we label
$\tau_{+}$ and $\tau_{-}$. Assuming the model is correct, the same value of $\tau_P$ applies for the 
$pp$ case and so we have enough information to determine $\tau(s)$ for $pp$.
\beqn
\tau_p=0.10,\, \tau_{+}= -0.51, \,\tau_{-}= 1.16
\eeqn
with the error ellipse for the two new couplings shown in Fig.~13.
\begin{figure}[thb]
\centerline{\epsfbox{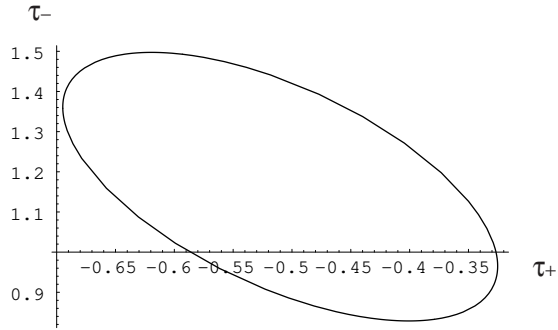}}
\medskip
{\caption[Delta]{Error ellipse for $pp$ Regge couplings with $C=\pm 1$.}}
\label{pm}
\end{figure}
These values are determined by the spin-flip couplings of $f + a_2$ and of $\omega +\rho$ respectively in a way to be made precise shortly. Because these are very different from $\tau_f$ and  $\tau_{\omega}$ we see  that the $a_2$ and $\rho$ couplings must be very large. This has been noticed from earlier data at lower energy and from $\pi^{\pm} N$ data \cite{berger}. We also see that these new couplings are not very well determined in spite of the very nice $pp$ data. This will surely improve in the future. On the other hand, the implied statistical error $\Delta A_N$ on the predicted $A_N$ is not so bad. See Fig.~14.
\begin{figure}[h]
\centerline{\epsfbox{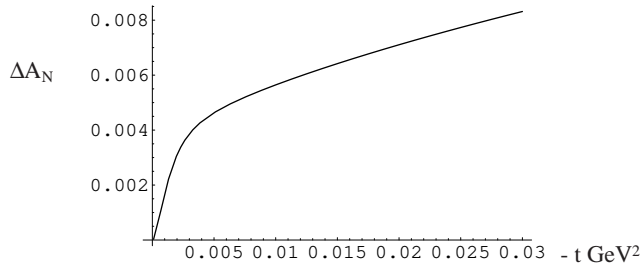}}
\medskip
{\caption[Delta] { The $1\sigma$ error on $A_N$ implied by the errors on $\tau$ as determined in this section.}}
\label{A_N error}
\end{figure}

We used these parameters to predict  $A_N$ at 24 {\rm GeV} before it was measured. The comparison to data is shown in Fig.~14:
\begin{figure}[b]
\centerline{\epsfbox{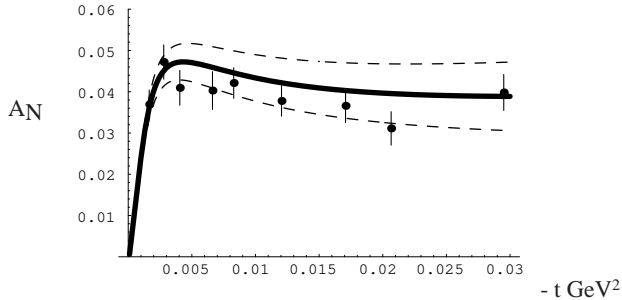}}
\medskip
{\caption[Delta] { Data for $A_N$ at $p_L=24 \,{\rm GeV}/c $ and prediction using parameters determined from 100~\,{\rm GeV}/c $pp$ measurement in conjunction with $\tau_P$ from $pC$. Dashed lines are $1\sigma$ errors.}}
\label{A_N 24}
\end{figure}
This shows the model is not bad: good enough to use for 10\%  polarimetry but not for 5\%.
\subsubsection{Joint fit to $\tau$'s for  $pp$ at 24 and 100 {\rm GeV}/c}
We use the measured values with known polarization to determine the three unknown $pp$ spin-flip factors, and determine the $\tau$'s that minimize the sum of the chi-square functions using the measured correlation errors. The results of this procedure is
\beqn
\tau_P= .068 \pm 0.054, \tau_{+}=-0.444 \pm 0.443, \tau_{-}=0.897 \pm  0.611
\eeqn
These values are consistent with the values using the previous method, but the errors are very large.  We recall that the most accurate value of $\tau_P= 0.10 \pm 0.01$ was obtained by the $pC$ analysis. Some have imagined  that $\tau(s)=0$; this is consistent with the 100 {\rm GeV} data, but at 24 {\rm GeV} it has a chi-square of 35.5 \cite{okada2}.  It is difficult to do better than this with the small values of $\tau$ with large uncertainties. Other attempts were made: for example by fitting the shape of the  $pp$  data at 24 {\rm GeV} with $\tau(100)$ together or by jointly fitting  $pp$ and $pC$. Somewhat different values of the Regge spin-flip factors are found but always within the errors which were found to be even larger.

\subsection{Predictions for higher energy}
\subsubsection{250 {\rm GeV}}
It is expected that there will be a run with polarized protons at 250 {\rm GeV}/c in the near future so we present here the prediction using the value for $\tau$  given in Section 4.2.1.
\begin{figure}[thb]
\centerline{\epsfbox{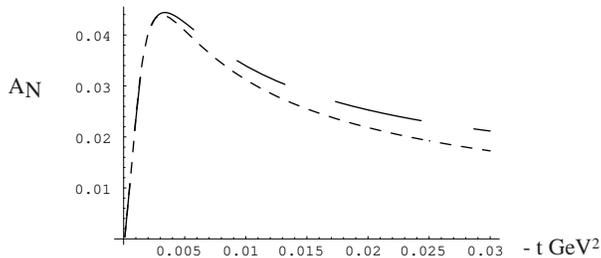}}
\medskip
{\caption[Delta] { Prediction for $A_N$ at $p_L=250 {\rm GeV}/c $ (long dashes) compared to fit at 100 {\rm GeV}/c (short dashes)}}
\label{250 and 100}
\end{figure}
Although $\tau$ remains small at 250 {\rm GeV}/c, the change from 100 {\rm GeV}/c is relatively large. The predicted energy dependence of $\tau$ over the RHIC fixed target range is shown in Fig.~17.
\begin{figure*}[thb]
\centerline{\epsfbox{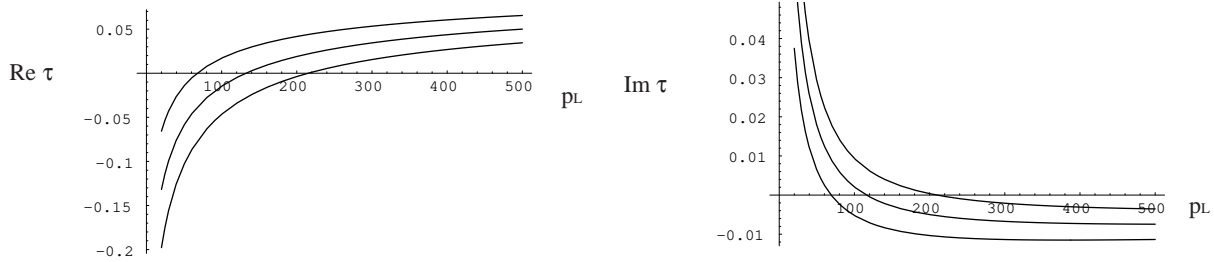}}
\medskip
{\caption[Delta] { Energy dependence of  ${\rm Re}\,\tau$ and ${\rm Im}\,\tau$ with errors through the RHIC fixed target energy range.}}
\label{pp energy dep.}
\end{figure*}
\subsubsection{pp2pp at $s=200^2$.}
There will soon be runs with colliding beams of polarized protons at various energies. The first one has already taken place with 100 {\rm GeV}/c beams in both rings. The prediction  of our model with the parameters just determined is
$\tau(200^2) =( 0.093 \pm  0.012) +i(-0.002 \pm 0.0008)$.

In 2005 the results of the first measurement of $A_N$ in the collider mode at RHIC were reported \cite{wlodek}. The spin flip parameter $r_5$ was determined with large errors to be
$r_5=(-0.033 \pm 0.035) + i(-0.43\pm 0.56) $ with the correlation being such that a zero value for both lies within the $1\sigma$ ellipse. To compare with our prediction we convert our prediction for $\tau$ to a prediction for $r_5$ using the calculated $\rho_{pp}(200^2)=0.127$ :
\beq
r_{5 \,prediction}=0.0138 \pm 0.008 +i(0.093 \pm 0.012)
\eeq
The comparison is shown in Fig.~18 where we have calculated the error ellipse for $r_5$ from \cite{wlodek};  the prediction, shown as a black ball to indicate approximate errors, is a little bit outside the $1\sigma$ ellipse.
\begin{figure}[thb]
\centerline{\epsfbox{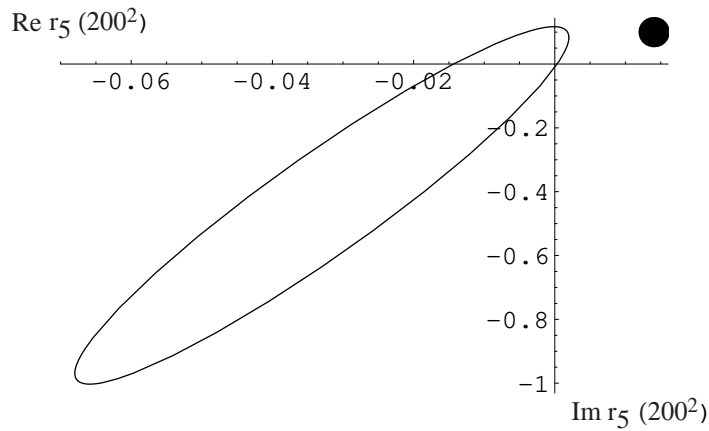}}
\medskip
{\caption{ Comparison of the value of $r_5$ with errors from $pp2pp$ with the model prediction with estimated errors marked by a black ball.}}
\label{Ann}
\end{figure}

This is a very big step in energy, and so we are happy that the measurement and our model prediction are compatible.  However, given the small errors of the model's predicton, more accurate $A_N$ measurement is needed at high energy. This is expected to be achieved in the near future.
\section{Regge couplings}
We would like to use these results to calculate $A_{NN}$, for example. In order to predict $A_{NN}$ we need to extract the factorized residues. There is not quite enough data to do this. The pomeron coupling is completely fixed but the $Y_1$ term in $g(s) $ gives $\beta_F^2+\beta_{a2}^2$, $\tau_{+} Y_1$ gives $(\beta_F^f \beta_{F} + \beta_{a2}^f \beta_{a2})$, $\tau_{f} Y_f$ gives $\beta_{F}^f \beta_F$. Similar for $C=-1$ poles: $Y_2$ gives $\beta_{\rho}^2 +\beta_{\omega}^2$, $\tau_{-} Y_2$ give $( \beta_{\rho}^f \beta_{\rho} + \beta_{\omega}^f \beta_{\omega})$ and $\tau_{\omega} Y_{\omega}$ gives $\beta_{\omega}^f \beta_{\omega}$.  We still need to pull out the $\rho$ and $ a2$ non-flip couplings.  We use $(Y_2-Y_{\omega})/2$ to give us $ \beta_{\rho}^2 $  and $(Y_1-Y_f)/2$ to give us $ \beta_{a2}^2$ so everything is determined. (Since there is not nearly as much data for $pn$ and the energy range is much smaller than for $pp$, there must be some hidden errors in this parametrization). 

What to use for Pomeron?  It doesn't have standard form here, but it doesn't matter for our program because its flip-factor $\tau_P$   is determined from the fit to $pC $ between $p=24$ and $p=100$. We will assume that, as with a factorized Regge pole, that the double flip pomeron will have $\tau_{P}^2(-t/m^2)$ before absorption. 

In this section we will use the central values that were given (with their errors) in the preceding sections.
\beqn
\tau_P&=& 0.10, \\ \nonumber
\tau_f &=& -0.79, \\ \nonumber
\tau_{\omega}&=& 0.516, \\ \nonumber
\tau_{+}&=& -0.510, \\ \nonumber
\tau_{-}&=& 1.163.
\eeqn
We use these values plus $Y_1,Y_2, Y_f, Y_{\omega}$ to produce the table
\begin{table}[h]
\centering
$\begin{array}{|c|c|c|}
\hline
\mbox{coupling} & \mbox{value}& \mbox{error} \\ 
\hline
\beta_f & 10.312 & 0.100\\
\beta_{\omega}& 9.027 & 0.035 \\
\beta_{a2} &1.774& 0.582 \\ 
\beta_{\rho} & 2.09& 0.150 \\ 
\beta_f^f & -8.147& 0.518 \\
\beta_{\omega}^f & 4.676  & 0.591 \\
\beta_{a2}^f & 15.877 & 12.729 \\ 
\beta_{\rho}^f & 27.652 & 13.424
\\ \hline 
\end{array}$
\caption{The non-flip and flip residues for the pure Regge pole model as normalized in the section}
\end{table}

There are a few points to be made here: although we have not attempted to use the same normalization as earlier works \cite{berger, irving} we can see similar patterns here: the $f$ and $\omega$ non-flip couplings are approximately the same as are the $\rho$ and $a2$; the $I=0$ non-flip are much larger than the $I=1$ but the $I=1$ flip are very much larger than the $I=0$ flip couplings. Unfortunately, the errors here for $I=1$ flip are very large. There are two reasons for this large error: (1) the ratio for flip to non-flip for $I=1$ has fairly large errors, about $30\%$, see Fig.~13, and (2) both $I=1$ non-flip couplings are rather small amplifying these errors in the calculation using factorization of the spin-flip factors into flip times non-flip residues. These errors will undermine our attempts to calculate the double-spin flip amplitude $A_{NN}$ which is based on this factorization.
\section{Double transverse spin asymmetry}
The natural next step to take in applying this model is to see what it says about $A_{NN}$. During the recent RHIC data taking some measurements of this quantity have been made in $pp$ scattering at three energies: 24 {\rm GeV}/c and 100 {\rm GeV}/c protons on a fixed hydrogen gas jet target \cite{okada2} and in colliding 100 {\rm GeV} beams {\cite{wlodek2}}. The asymmetry is calculated using Eq.~\ref{eq:asymdef}. Note that near $t=0$ $A_{NN}$ and $A_{SS}$ become nearly equal and in \cite{wlodek2} they measure both.

We will write in analogy to Eq.~\ref{phi5} and \cite{berger,irving}
\begin{widetext}
\beq
\phi_2(s,t)= \f {1}{8 \pi} \{-\tau_P^2 \, s\, gP(s) + ( \left. \beta_f^ f \right. ^2+  \left. \beta_{a2}^f \right.^2) \f{1 +\exp(-i \pi \alpha_f)}{\sin \pi \alpha_f} s^{\alpha_f} + ( \left. \beta_{\omega}^ f \right. ^2+  \left. \beta_{\rho}^ f \right. ^2) \f{1 -\exp(-i \pi \alpha_\omega)}{\sin \pi \alpha_{\omega}}s^{\alpha_{\omega}} \} \f{(-t)}{m^2}
\label{phi2}
\eeq
\end{widetext}
This is the form given by the pure Regge pole model. The physics will require that this be modified soon.  First, parity conservation of the hadronic interaction and factorization of the poles requires that $\phi_2(s,t) =-\phi_4(s,t)$ \cite{BKLST} . At the same time angular momentum conservation requires that  $\phi_4(s,t)$ vanishes with $t$ as $t \rightarrow 0$. This forces $\phi_2(s,t)$ to vanish in the same way and so kinematically suppress $A_{NN}$. Because the poles are suppressed, they cannot be the leading contribution near $t=0$ and so we must include Regge cut contributions. Because the cuts break factorization, the equality of $\phi_2$ and $-\phi_4$ is broken and the suppression is lost.

Of course, bringing  in cuts takes the model dependence of this calculation to a new level, so we will adopt the simplest possible approach to estimating cut effects, the ``absorptive cut" model of \cite{henyey}. This has been widely used in unpolarized calculations with some success, so we can hope that a reasonable estimate is obtained in this way, but claim nothing more than that. 

We will here follow the general ideas of Kane and Seidl \cite{KS} for absorption corrections to Regge poles. There is assumed to be one absorption factor $S(b)$  in $b$-space (impact parameter) for all amplitudes. It will be small for $b=0$ and rise to near 1 as $b \rightarrow 10 \,{\rm GeV}^{-1}$. It is derived from elastic rescattering plus an artfully constructed contribution from inelastic scattering. I will not try to model that but simply construct a simple form for $S$ that looks similar to what they calculate. We will need to correct most importantly the double-flip amplitude, but we will also need to calulate the effect on non-flip and single-flip. This will change the relation between the Cudell parameters and $\tau (s)$, and the Regge residues.  We will continue to assume that all terms have the same $e^{B(s) t/2}$ behaviour. 

We will assume here that $S$ is real to avoid too many parameters. So define
\beq
S_K(b/R_a) = 1 - K\, e^{-\f{b^2}{R_a^2}}
\eeq
$K$ is a real number that determines the strength of the absorption, and we will take $R_a^2=2 B(s)$. We will start from $\phi_1$  and $\phi_5$, transform the simple pole forms  to $b$-space, then convolute with $S$. This will change the relation between the Regge residues and the Cudell parameters as determined in $\phi_1 $ and $\phi_5$.  We treat the Pomeron slightly differently from the other Regge poles: since we began with a Pomeron which is manifestly  not a Regge pole, we assume that it already has the absorption cuts taken into account in both $\phi_1$ and $\phi_5$. In addition we assume that its spin dependence factorizes so the form is exactly as in Eq.~\ref{phi2} with the $\tau_P$ as determined from the $A_N$ fits.  We will plug these new parameters into $\phi_2 $ and then run it through the absorption machinary. The most important result of this will be to replace the $-t$ factor in $\phi_2 $ by a constant times $1/B(s)$.

We have used two different values of $K$: $K=1$ corresponding to complete absorption and $K=0.6$, the value favored by  \cite{KS}. $K$ could in principle depend on $s$ but the level of accuracy of our other parameters and of the current data does not really allow a serious consideration of this.The result of this simple model is that now the non-flip residues are increased by a factor of $1/\sqrt{0.5}$ for $K=1$ or $1/\sqrt{0.7}$ for $K=0.6$ from the values determined by the Cudell parameters. At the same time the spin-flip residues are  {\em decreased} by a factor of 0.94 or 0.98 for $K=1$ or $K=0.6$, respectively.  The most important change is the behaviour of $\phi_2$ as $t \rightarrow 0$; it now no longer vanishes but, in the small $t$ region, $-t$ is replaced by $\f{K}{2 B(s)}+ \mathcal{O}(t)$. The same calculation yields
\beq
\phi_4^{cut}(s, t) = B(s)\,t/8\, \phi_2^{cut}(s, z)
\eeq
explicitly breaking the factorization.
\begin{figure*}[thb]
\centerline{\epsfbox{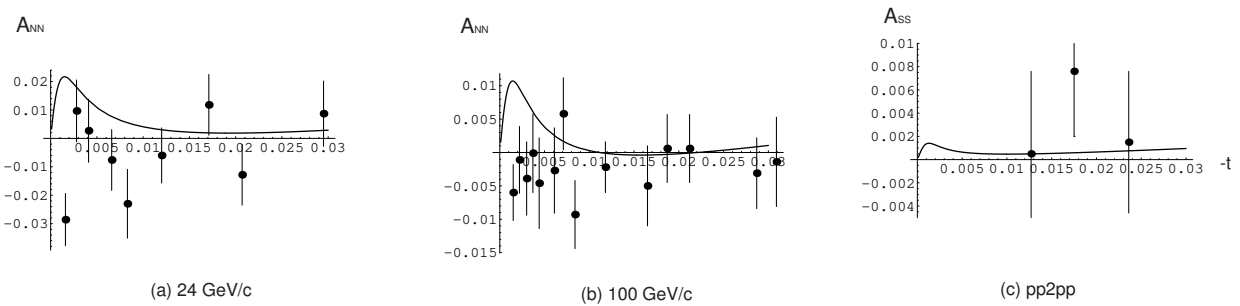}}
\medskip
{\caption{ Model prediction of $A_{NN}$ for $24\, {\rm GeV}/c$ and $100\, {\rm GeV}/c$ protons on gas jet target for absorption parameter $K=1.0$ with data from \cite{okada2} and $A_{SS}$ data from $pp2pp$ \cite{wlodek2}. }}
\label{Ann}
\end{figure*}
The model now predicts a significant value for $A_{NN}$ which we show in Fig.~19 for the value $K=1.0$. For $K=0.6$ the peaks in each case are about $0.6$ as high as in Fig.~19. It is clear from this that determination of $K$ or a validation of the model is not yet possible. (Actually the $pp2pp$  data here is for $A_{SS}$ but there is little difference between the two predictions.)
A convenient way of parametrizing the strength of the double-flip amplitude is through \cite{BKLST}
\beq
r_2(s)= \f{\phi_2(s,0)}{2 \,{\rm Im}\, \phi_1(s,0)}.
\eeq
Note that the transverse total cross section asymmetry $\Delta \sigma_{T} = - 2\, {\rm Im} r_2 \, \sigma_{tot}$. $r_2$ is plotted over the fixed target range in Fig.~20 along with $1\sigma$ errors. 
(For $K=0.6$ the value of $r_2$ is about $0.6$ times the magnitude.)
Its very small value at $s=200^2 \,{\rm GeV}^2$  is $ -0.0005 + 0.0002i$, far smaller than the errors of the recent experiment \cite{wlodek2}. This will be a challenge to observe.
\begin{figure*}[!]
\centerline{\epsfbox{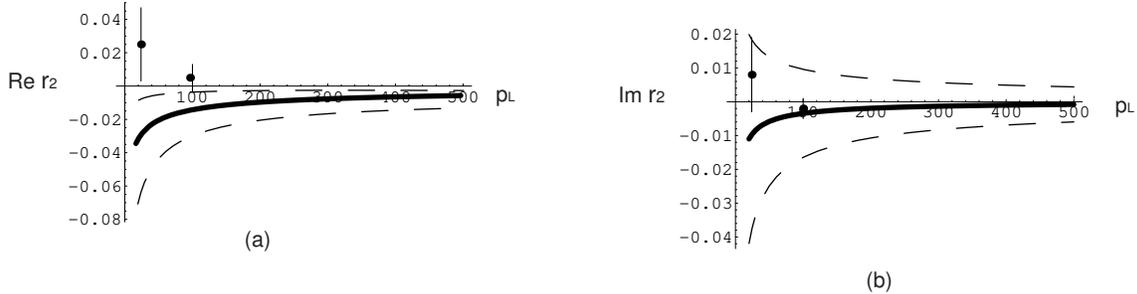}}
\medskip
{\caption{Model prediction of $r_2(s)$ parameter $K=1.0$ with data from \cite{okada2}. }}
\label{r2}
\end{figure*}
\subsection{The Odderon}
Several years ago Elliot Leader and the author proposed using $A_{NN}$ as a means for searching for the Odderon \cite{LT}. The idea behind this was that, because the Odderon is odd-signature it would be asymptotically real in contrast to the Pomeron \cite{pdb}. The formula for $A_{NN}$, Eq.~2.4, contains $Re(\phi_2^* \phi_1)$ and there the signal would be enhanced at small $t$ due to Coulomb-nuclear interference. This gives a very characteristic signature that we proposed using to look for the Odderon. Since at that time we had no information at all about the spin flip couplings, our discussion was purely qualitative. Now we have a limited amount of information from the preceding discusson and can see if there is any sense to our proposal. 

So we simply add to the expression for $\phi_2$ a simple Odderon pole at $J=0.96$ as a good guess based on QCD from Nicolescu \cite{Nicolescu},  a spin-flip coupling of unknown value $\beta_O$ and zero non-flip coupling since there is no sign of it in unpolarized experiments \cite{Ewarz}.  We then process the new $\phi_2$ through the absorption model and predict $A_{NN}$. 

Since we have no idea what the Odderon spin-flip coupling should be, we show in Fig.~21 the prediction using $K=0.6$ for four values of $\beta_O$, (0,1,2,3), for the three energies where there is data. (For $K=1$ the prediction is about a factor of 5/3 larger.) From this we conclude that we cannot rule out an Odderon with a modest spin-flip coupling, but it clearly must be smaller than the normal Regge flip couplings. (Recall that $\beta_{\omega}^f=4.67$ and $\beta_{\rho}^f=27.65$ for comparison.)
\begin{figure*}[thb!]
\centerline{\epsfbox{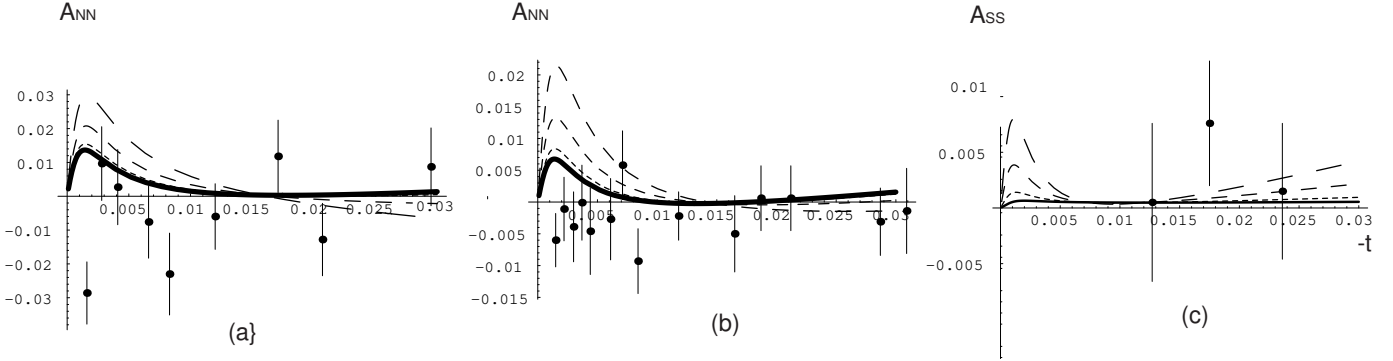}}
\medskip
{\caption{ Prediction for $A_{NN}$ or $A_{SS}$ with $K=0.6$ at (a) $p_L=24\,{\rm GeV}/c , \,(b)\, p_L=100\,{\rm GeV}/c, \newline (c)\, s=200^2 {\rm GeV}^2$ with data from \cite{okada2,wlodek} with Odderon coupling $\beta_O=0$(solid),1(short), 2(medium) and 3(long) .}}
\label{Ann}
\end{figure*}
\section{summary}
This paper serves as a report on the large amount of work done in conjunction with the polarized proton program at RHIC. In addition to providing a model which works at the 10\% level for the energy dependence of the the analyzing power of elastic scattering needed for proton polarimetry, it also provides some useful information about the spin dependence of dominant Regge poles. Most notably, the data indicate that the Pomeron has a small but significant spin-flip coupling. These results  allow the exploration of the double spin flip asymmetry $A_{NN}$ for which some data over a wide energy range are now available \cite{okada2,wlodek2}, along with a concrete realization of a proposed Odderon search \cite{LT}. Our results are limited by the rather large errors on some of the data we have used and we expect that this will be remedied in the not too distant future, leading to more precise polarimetry and better determination of the Regge spin-flip couplings and so better values for $A_{NN}$.
\newpage
\begin{acknowledgments}
Many people had important influence on this work; in particular my many experimental colleagues who initiated and maintained my interest in the topic by providing a steady stream of new data and questions concerning it over many years: Yousef Makdisi, Gerry Bunce, Dimitri Svirida,  Wlodek Guryn, Osamu Jinnouchi, Alesandro Bravar, and Hiromi Okada most extensively. My theoretical colleagues who I worked with on several related problems also deepened my understanding: Elliot Leader, Nigel Buttimore, Boris Kopeliovich and Jacques Soffer, and my old friends from Ann Arbor, Alan Krisch and Gordy Kane.
\end{acknowledgments}

\end{document}